# A Data-driven Framework for Error Estimation and Mesh-Model Optimization in System-level Thermal-Hydraulic Simulation

Han Bao, Nam Dinh, Jeffrey Lane, Robert Youngblood

## ABSTRACT

Over the past decades, several computer codes have been developed for simulation and analysis of thermal-hydraulics and system response in nuclear reactors under operating, abnormal transient, and accident conditions. However, simulation errors and uncertainties still inevitably exist even while these codes have been extensively assessed and used. In this work, a data-driven framework (Optimal Mesh/Model Information System, OMIS) is formulated and demonstrated to estimate simulation error and suggest optimal selection of computational mesh size (i.e., nodalization) and constitutive correlations (e.g., wall functions and turbulence models) for low-fidelity, coarse-mesh thermal-hydraulic simulation, in order to achieve accuracy comparable to that of high-fidelity simulation. Using results from high-fidelity simulations and experimental data with many fast-running low-fidelity simulations, an error database is built and used to train a machine learning model that can determine the relationship between local simulation error and local physical features. This machine learning model is then used to generate insight and help correct low-fidelity simulations for similar physical conditions. The OMIS framework is designed as a modularized six-step procedure and accomplished with state-of-the-art methods and algorithms. A mixed-convection case study was performed to illustrate the entire framework.

# ACRONYMS

| | | | |
|---|---|---|---|
| BWR | Boiling Water Reactor | ML | Machine Learning |
| CFD | Computational Fluid Dynamics | MSE | Mean Squared Error |
| CG-CFD | Coarse Grid CFD | NN | Neural Network |
| DNN | Deep Neural Network | NPP | Nuclear Power Plant |
| DNS | Direct Numerical Simulation | OMIS | Optimal Mesh/Model Information System |
| EMDAP | Evaluation Model Development and Assessment Process | OOB | Out-Of-Bag |
| FDM | Finite Difference Method | PCC | Physics Coverage Condition |
| FEM | Finite Element Method | PDE | Partial Differential Equation |
| FNN | Feedforward Neural Network | PDF | Probability Density Function |
| GELE | Global Extrapolation through Local Extrapolation | PF | Physical Feature |
| GELI | Global Extrapolation through Local Interpolation | PFC | Physical Feature Coverage |
| GEP | Gene Expression Programming | PIRT | Phenomena Identification and Ranking Table |
| GILE | Global Interpolation through Local Extrapolation | PVIM | Permutation Variable Importance Measure |
| GILI | Global Interpolation through Local Interpolation | QoI | Quantity of Interest |
| GPR | Gaussian Process Regression | RANS | Reynolds-averaged Navier–Stokes |
| HF | High-Fidelity | REMDAP | Risk-informed EMDAP |
| HL | Hidden Layer | RFR | Random Forest Regression |
| IC/BC | Initial Condition/Boundary Condition | RISMC | Risk-Informed Safety Margin Characterization |
| KDE | Kernel Density Estimation | t-SNE | t-Distributed Stochastic Neighbor Embedding |
| LF | Low-Fidelity | V&V | Verification and Validation |
| LOCA | Loss-of Coolant Accident | | |

# NOMENCLATURE

*Arabic*

| | |
|---|---|
| $C_p$ | Specific heat |
| $D_h$ | Hydraulic diameter |
| $d_{KDE}$ | KDE distance |
| $d_{Ma}$ | Mahalanobis distance |
| $g$ | Gravitational acceleration |
| $h$ | Heat transfer coefficient |
| $H$ | Height |
| $l$ | Characteristic length |
| $P$ | Pressure |
| $T$ | Temperature |
| $p_{KDE}$ | KDE probability |
| $U$ | Velocity |
| $w$ | Width of the cell |
| $y$ | Distance to the wall |
| $Re$ | Reynolds Number |
| $Gr$ | Grashof Number |
| $Pr$ | Prandtl Number |
| $Ri$ | Richardson Number |

*Greek Symbols*

| | |
|---|---|
| $k$ | Kinetic energy |
| $\mu$ | Dynamic viscosity |
| $\nu$ | Kinetic viscosity |
| $\varepsilon$ | Dissipation rate |
| $\varepsilon_{mesh}$ | Mesh-induced error |
| $\varepsilon_{model}$ | Model error |
| $\varepsilon_{sim}$ | Simulation error |
| $\rho$ | Density |
| $\lambda$ | Thermal conductivity |

*Subscripts*

| | |
|---|---|
| $T$ | Turbulent |
| $l$ | Liquid |
| $v$ | Vapor |
| $f$ | Free, no contact to wall |
| $w$ | Wall |
| $o$ | Outlet |

1. **INTRODUCTION**

1.1. **Motivation**

Quantification of Nuclear Power Plant (NPP) safety risk requires a systematic and yet practical approach to identification of accident scenarios, assessment of their likelihood and consequences. Such an approach is provided by the Risk-Informed Safety Margin Characterization (RISMC) framework [1], whose realization requires computationally robust and affordable methods for sufficiently accurate simulation of complex multi-dimensional physical phenomena, such as turbulence, heat transfer, and multi-phase flow. While several types of codes have been extensively developed, assessed, and used to support design, licensing, and safety analysis of the plants, simulation errors still inevitably exist.

A significant difficulty is caused by the complexity of these multi-dimensional multi-phase physical phenomena in the transient scenarios. These phenomena occur in the different NPP components with complex geometries and structures, making it impossible to perfectly model and simulate the entire NPP thermal-hydraulic systems in all time and length scales. For present purposes, it is useful to consider three types of computational codes used for thermal-hydraulic analysis. The first type is called lumped-parameter code or system code, such as RELAP 5 and TRAC. These codes describe an NPP thermal-hydraulic system as a network of simple control volumes connected with junctions. Turbulence effects are not directly modeled but, up to a point, can be considered using assumed flow-loss coefficients in the momentum equation. [2] When time and geometry averaging approaches are applied on the local instantaneous two-fluid models, in an effort to speed up computation, much local information is lost. The second type is Computational Fluid Dynamics (CFD), which has become commonly used for solving transport equations of fluid mechanics (continuity, momentum and energy) using a local instantaneous formulation. These CFD codes (e.g., STAR-CCM+) consider turbulence effects using different turbulent models. System-level thermal-hydraulic analysis using CFD codes is computationally expensive, since a million cells might be needed even for modeling of a single NPP component. The third type of code is a coarse-mesh, CFD-like code, such as GOTHIC. [3] These codes provide a 3D simulation capability, use a coarse mesh size with, the sub-grid phenomena in boundary layer being well captured by adequate constitutive correlations (e.g., wall functions and turbulence models), which makes these codes very computationally efficient. Different from standard system codes (with much loss of local information) and standard CFD codes (with huge computational cost), these codes have natural advantages for achieving sufficient accuracy for long-term multiple-component system-level simulation. These CFD-like codes have been extensively used for containment thermal-hydraulic analysis. [4-8]

However, two main error sources exist in the application of these coarse-mesh CFD-like codes. One is the model error due to physical simplification and mathematical approximation on these applied models, correlations, and assumptions. They solve the integral form of conservation equations for mass, momentum, and energy for multi-component, multi-phase flow. Boundary-layer correlations are applied for heat, mass, and momentum exchanges between the fluid and the structures, rather than attempting to resolve the boundary layers specifically. The respective characteristic lengths of these empirical correlations are calculated by default using the local mesh size. Therefore, the mesh size greatly affects the performance of the empirical correlations in the local near-wall cells and becomes a key model parameter that determines whether the correlations are being applied in appropriate ranges. Another source of error is mesh-induced error, which indicates the information loss of conservative and constitutive equations during the application of time and space averaging approaches. The local instantaneous Partial Differential Equations (PDEs) for mass, momentum, and energy balance are space-averaged to obtain the finite volume equations. Simulation results represent the averaged values of parameters over specified regions, which ignores the local gradient information. A similar concept, discretization error, is proposed from the classic Verification and Validation (V&V) point of view for the solving of PDEs, which assumes that when mesh size goes to zero the solution of PDEs converges. However, due to the correlation-based design in the simplified boundary-layer treatment, these CFD-like codes (e.g., GOTHIC) are not expected to converge when mesh size goes to zero, because these empirical correlations may no longer be valid for very fine mesh. Taking GOTHIC as an example, it applies finite volume technique with cell volume and surface porosities for complex geometries. The local instantaneous PDEs for mass, momentum and energy are time and space averaged to obtain the finite volume equations. Results from GOTHIC represent the averaged values of parameters over specified regions, not the exact value at the central points of the regions. Other numerical errors have less influence on the simulation compared to model error and mesh-induced error.

Since both main error sources are tightly connected with local mesh size, the nodalization of control volumes determines whether the user can get a relatively good simulation result. The finite mesh/volume approach, particularly in the coarse scheme of NPP simulations, could fail to capture the expected local behaviors of the fluids (sharp gradients of variables) due to limited resolution. On the other hand, a finer nodalization could introduce an improper extrapolation of boundary-layer empirical correlations. All these factors make the selection of mesh size and model information (model parameter and model form) an important but tricky task in the system-level thermal-hydraulic modeling and simulation using these CFD-like codes. Generally, the mesh size and models are selected based on previous simulation experience or available in modeling guidelines derived from past benchmarking and application of the code; however, this kind of "educated guess" or engineering judgement may lead to an unknown error for the new physical conditions. Therefore, a systematic approach is needed to provide optimal selections of models and coarse mesh size.

## 1.2. Objective of This Work

In order to provide error prediction and advice on the selections of optimal mesh size and models for system-level thermal hydraulic simulation, a data-driven framework (Optimal Mesh/Model Information System, OMIS) is proposed in this work. The OMIS framework is developed for thermal-hydraulic codes that have the following features: they use coarse mesh sizes and apply simplified boundary-layer correlations whose applicable ranges depend on respective characteristic lengths, such as CFD-like codes or coarse-mesh Reynolds-averaged Navier–Stokes (RANS) methods with wall functions. This coarse-mesh framework benefits from the application of machine learning algorithms and the computational efficiency of coarse-mesh codes for system-level thermal-hydraulic modeling and simulation.

Over the past few decades, many concepts of nuclear reactor have been proposed with different components, geometries, and powers. The respective global physical conditions might be an "extrapolation" from previous designs or simulations, which brings large uncertainty into the application. The relevant thermal-hydraulic experiments with a wide range of scale and structures must be designed for respective code development and validation. This extrapolation may limit the applicability of the global response or the corresponding macroscale data; however, local physics such as the interaction between liquid, vapor and heat structure may not change. This makes it possible that local physical parameters or variables in the local cells are similar even if the global physical condition totally changes. The Global Extrapolation through Local Interpolation (GELI) condition is defined to represent the situation where the global physical condition of target case is identified as an extrapolation of existing cases, but the local physics are similar. The extrapolation of global physics indicates different global physical conditions such as a set of characteristic non-dimensional parameters, or different initial conditions / boundary conditions (IC/BCs), or different geometries/structures, or dimensions. The underlying local physics can be represented by a set of Physical Features (PFs). The interpolation or similarity of local physics is dependent on the identification of PFs, data quality and quantity. The local similarity in GELI condition makes it feasible to derive great benefits from the existing data to estimate the target case. Instead of endlessly evaluating the applicable ranges of models and scaling uncertainty, exploring the similarity of local physics opens another door to overcome the scaling issues in global extrapolations. GELI is one of the Physics Coverage Conditions (PCCs); the other three are Global Interpolation through Local Interpolation (GILI), Global Interpolation through Local Extrapolation (GILE), and Global Extrapolation through Local Extrapolation (GELE). Targeting the GELI condition, the OMIS framework achieves a potential scalability to the globally extrapolated conditions by concentrating on the similarity of local physics.

Section 2 reviews the machine learning applications on thermal-hydraulic analysis. Section 3 describes the methodology of the proposed OMIS framework. Section 4 discusses the case study on mixed convection simulation. Section 5 summarizes the results.

## 2. MACHINE LEARNING APPLICATIONS IN THERMAL-HYDRAULIC ANALYSIS

### 2.1. Propose of Data-driven Modeling and Simulation Framework

In 2013, some new perspectives were proposed in nuclear reactor thermal-hydraulics. [9] A concept of "data-driven modeling and simulation framework" was proposed to enable the simulation code applying pattern recognition and statistical analysis to obtain required closure information directly from the relevant database generated from huge amounts of experiments and simulations. This concept makes direct use of existing rich high-fidelity data, instead of converting the data into separate physical models, causing a great deal of information to be abandoned. For conditions where directly applicable data is absent, the information can be predicted based on the near-by conditions included in the database. Uncertainty due to the lack of data can be reduced as new data becomes available. High-fidelity data refers to data that have been adequately validated and has a potential to be used to reduce the simulation uncertainty in low-fidelity modeling and simulation. Nowadays, the explosive development of machine learning algorithms and massive data available from numerical simulations make the idea of "data-driven" realistic and feasible. A validation and uncertainty quantification framework for Eulerian-Eulerian two-fluid-model based multiphase CFD solver has been formulated based on the integration of data and model. [10] The proposed framework applies a Bayesian method to inversely quantify the uncertainty of the solver predictions with the support of multiple experimental data. However, the numerical error introduced in the discretization of the PDEs is not considered. In the Integrated Research Project titled "Development and Application of a Data-Driven Methodology for Validation of Risk-Informed Safety Margin Characterization Models", a validation framework, named Risk-informed Evaluation Model Development and Assessment Process (REMDAP), is proposed for the validation of RISMC models, which is designed by combining data-driven and risk-informed concepts. [11]

## 2.2. Data-driven Modeling Applications on Fluid Dynamics

Many efforts have been made on the development of data-driven approaches in the study of fluid dynamics, especially the data-driven turbulence closures to deal with the issues from model form uncertainty and knowledge lack of turbulence. Early in 2002, Milano used Direct Numerical Simulation (DNS) results as high-fidelity data to train a Neural Network (NN) to replicate near-wall channel flows but did not apply these NNs on forward models for turbulent flow prediction. [12] Tracey and Duraisamy used NNs to predict the Reynolds stress anisotropy and source terms for turbulence transport equations. [13] Parish and Duraisamy introduced a multiplicative correction term for the turbulence transport equations using Gaussian Process Regression (GPR) with the uncertainty of this correction term quantified. [14] Zhang and Duraisamy also applied NNs to predict a correction factor for the turbulent production term in channel flow, which could affect the magnitude but not the anisotropy of the predicted Reynolds stress tensor. [15] Ling proposed the training of Random Forests (RFs) to predict the Reynolds stress anisotropy. [16] But Ling and Templeton also explored the capability of RFs and NNs in learning the invariance properties and concluded that RFs are limited in their ability to predict the full anisotropy tensor because they cannot easily enforce Galilean invariance for a tensor quantity. [17] So later Ling and Templeton used deep NNs with embedded invariance to predict the Reynolds stress anisotropy. [18] Different from those data-driven approaches above that directly predict Reynolds stress, Wang and Xiao proposed to apply RFs to predict the Reynolds stress discrepancy. [19] It should be noted that several well-selected physical features are used as the training input instead of physical coordinates in this approach. Another machine learning algorithm, Gene Expression Programming (GEP) was applied by Weatheritt and Sandberg to formulate the non-linear constitutive stress-strain relationships for turbulence modeling. [20] Recently, Zhu and Dinh performed a data-driven approach to model turbulence Reynolds stress leveraging the potential of massive DNS data. [21] The approach is validated by a turbulence flow validation case: a parallel plane quasi-steady state turbulence flow case. Most approaches focused on how to deal with model form uncertainty of RANS turbulence modeling without considering the numerical error due to discretization. Hanna and Dinh investigated the feasibility of a Coarse Grid CFD (CG-CFD) approach by utilizing machine learning algorithms to produce a surrogate model that predicts the CG-CFD local errors to correct the variables of interest. [22] This work focused on the correction of discretization error of CG-CFD without considering the model errors that may be introduced in applications of CFD to thermal-hydraulic analysis.

According to a classification of machine learning frameworks for thermal fluid simulation introduced by Chang et al. [23], most of these efforts mainly belong to framework Type I or Type II. Type I aims at developing new closure models by assuming that conservation equations and closure models are scale separable. Type II concentrates on reducing the uncertainty of low-fidelity simulation by "learning" from high-fidelity data. Both require a thorough understanding of the physical system and sufficient prior knowledge on closure models. These limitations make current data-driven approaches for specific local closure laws inapplicable to the complex system-level thermal-hydraulic application. The complexity of the prior knowledge that is needed increases very significantly when all the components, processes and involved phenomena in reactor systems need to be considered together. A data-driven approach with less knowledge required is urgently needed for complex situations, especially when a great amount of high-fidelity data and computation capability are available. Type V relies entirely on machine learning algorithms to discover the underlying physics directly from data, and does not require prior knowledge. The CG-CFD approach belongs to Type V which does not have requirement for prior knowledge. However, none of the data-driven approaches reviewed above are designed for CFD-like or coarse-mesh CFD codes. These efforts analyzed model error and mesh-induced error separately with the other fixed; this is inapplicable to the coarse-mesh methods where mesh size is treated as a model parameter and mesh convergence is not expected. To overcome this difficulty in the application of CFD-like codes, OMIS is developed to deal with these two error sources together, as shown in Figure 1. The OMIS framework is considered as a Type V framework since it treats the physical models, coarse mesh sizes and numerical solvers as an integrated model, which can be considered as a surrogate of governing equations and closure correlations of low-fidelity code. It does not need relevant prior knowledge, and purely depends on existing data. Besides, compared to current data-driven efforts, the OMIS framework is successfully applied in thermal-hydraulic analysis in this paper, not just in adiabatic fluid dynamics where previous efforts were focused. A demonstration study of OMIS predictive capability has been performed on a turbulent mixing case. [24]

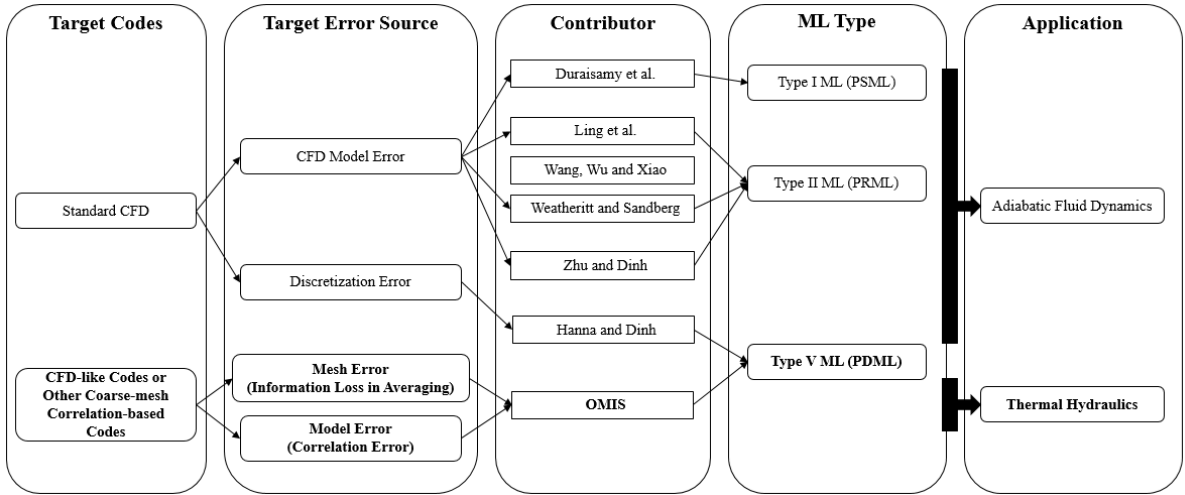

Figure 1. Review of Machine Learning Applications on Thermal-Hydraulic Modeling

## 3. METHODOLOGY

### 3.1. Hypotheses

The total simulation error ($\varepsilon_{sim}$) for the physics of interest using coarse-mesh CFD-like codes includes the model error ($\varepsilon_{model}$), mesh-induced error ($\varepsilon_{mesh}$) and other numerical errors having less effect on the total simulation error. By treating these two main error sources together, the OMIS framework develops a surrogate model to identify the relationship between $\varepsilon$ and specific local Physical Features (PFs), as shown in Figure 2. The identification of PFs integrates the physical information of the system of interest, model information and the effect of mesh size. Once the error function $\varepsilon_{sim} = f(PFs)$ is developed and evaluated via existing data and the application of machine learning algorithms, the simulation error for a new condition with the specific mesh and model is predictable. The claims that need to be established for this are: (1). Simulation error can be represented as a function of key Physical Features (PFs) which integrate the information from local physics, applied models and local mesh sizes; (2). "$\varepsilon_{sim} = f(PF)$" is not a fixed correlation, it represents the relationship between simulation errors and physical features, which is improvable when new qualified data are added into training data; (3). Similarity between training data and testing data determines the predictive capability of trained machine learning models for the test case.

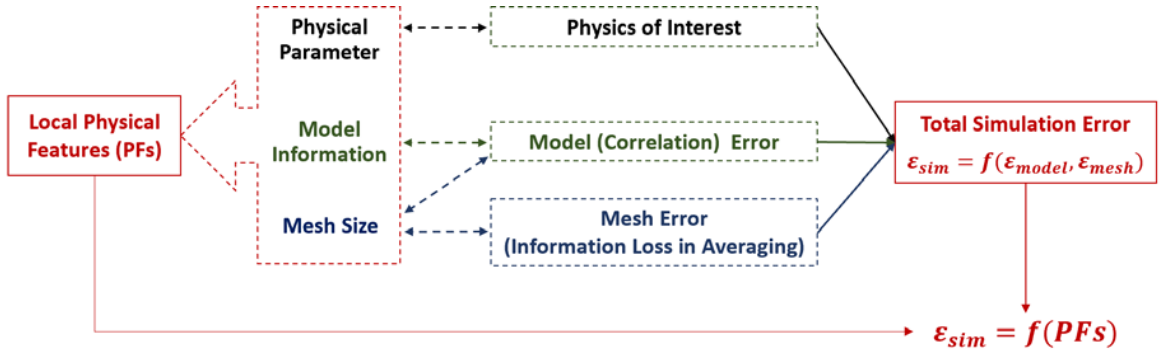

Figure 2. Central Idea of OMIS: Local Data Training for Error Estimation

### 3.2. Framework Formulation

This data-driven mesh-model optimization framework contains six independent steps as displayed in Figure 3.

#### 3.2.1. Step 1: Target Simulation Analysis

The items that should be specified in this step are (1) Key phenomena and global Quantities of Interest (QoIs) in the target case; (2) Applicable physical models for these key phenomena in the simulation tool; (3) Global parameters that represent the global physical condition of the target case; (4) Potential mesh sizes for specific control volumes. A PIRT (Phenomena Identification and Ranking Table) procedure should be executed to decompose the complex physics and identify the key phenomena in the target simulation. QoIs in system-level thermal-hydraulic simulation are normally global parameters and depend on the phenomena in the given scenario. These quantities represent the system behaviors of NPPs and provide information for decision-making. For example, in normal operation of Boiling Water Reactors (BWRs), the main steam line temperature and reactor vessel pressure are considered to be QoIs, since they reflect the performance of the heat removal of fission in the fuel bundles. The temperature/pressure in the wetwell and the hydrogen fraction in the drywell can be considered as the key QoIs if severe events happen, as in the Fukushima accident. The values of these QoIs could help the operators estimate the benefit and risk associated with the timing of injection of seawater into the units. Optimizing the prediction accuracy of these key QoIs is the purpose of the OMIS framework.

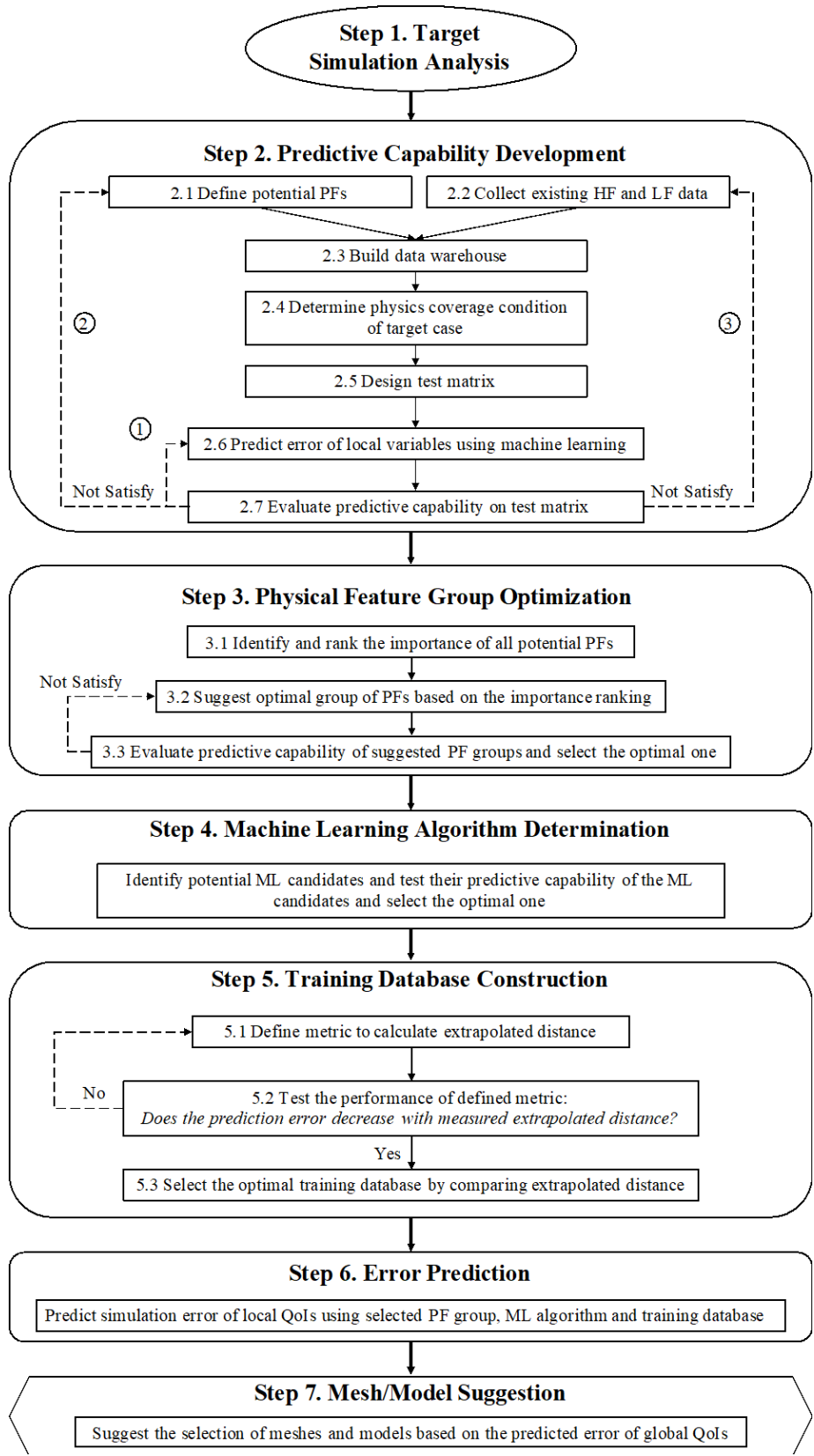

Figure 3. Diagram of the Optimal Mesh/Model Information System (OMIS) Framework

Some global parameters should be identified to represent the global physical condition of the target case, which helps the selection of data warehouse and the construction of test matrix. For example, for piping flow, the Reynolds (Re) number is identified as the key global physical parameter. Then the high-fidelity data with similar values of Re number should be collected into the data warehouse in Step 2. Based on Re number values in the target case and the training case, it should be determined which PCC condition the target belongs to. Lastly, according to the geometry and structure of the control volumes in the target case, a set of potential mesh sizes should be selected for different control volumes. Based on the capability of the simulation tool, these mesh sizes should be in an appropriate range where they are neither too fine, incurring too much computation cost, nor too coarse, losing too much local information.

### 3.2.2. Step 2: Predictive Capability Development

This step is to establish the predictive capability by preliminarily defining potential PFs, building the data warehouse, and evaluating the predictive capability of the selected database on test matrix.

- **Step 2.1: Define potential physical features**

The identification of PFs is guided by the physics decomposition and model evaluation executed in Step 1. To take physics scalability and regional information into consideration, the PF group includes the gradients of local variables and the local physical parameters that can represent the crucial local physical behaviors or closure relationships. All potential PFs that satisfy the definition should be considered and included in the initial selection of the PF group. Information of the physical system, model information and the effect of mesh size are ensured to be included and well represented in the PF group, as illustrated in Figure 4. The gradients of variables include 1-order and 2-order derivatives of variables calculated using central-difference formulas. They contain the regional (or local surrounding) information that represents the regional physical patterns. As displayed in Figure 5, the regional information obtained from the training dataset (as Case A) can be used to teach and inform the prediction of new conditions (as Case B) in GELI condition: if the regional information in part of Case A is similar to the part of Case B. More regional information may be involved if higher order derivatives are added into the local PF group.

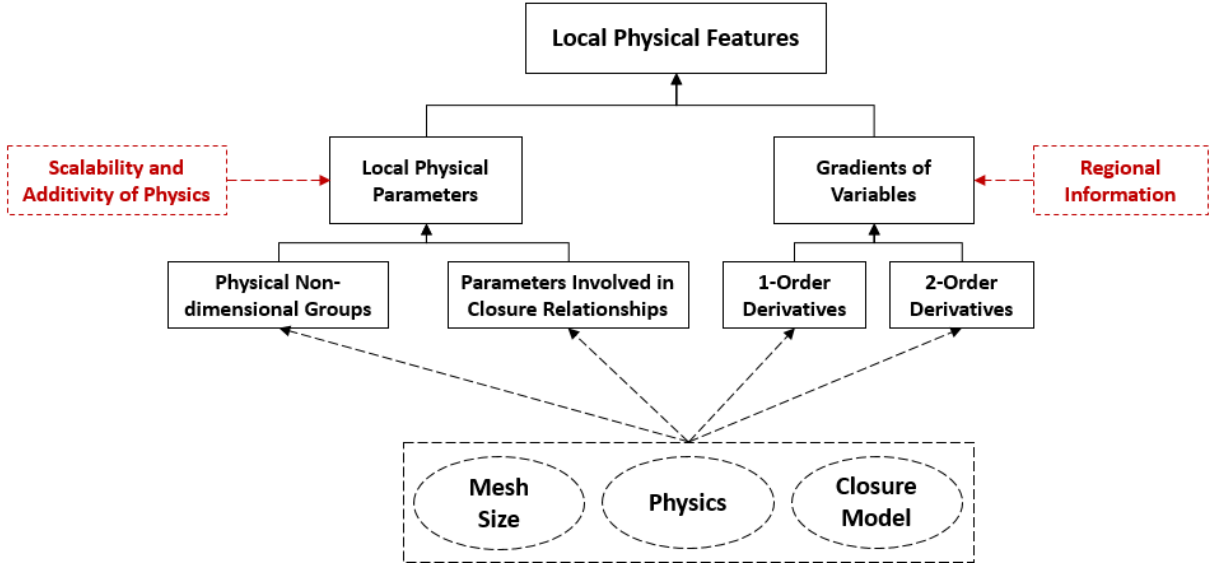

Figure 4. Identification and classification of Physical Feature

$$\left.\frac{\partial V}{\partial x_i}\right|_{(i,j)} = \frac{V_{i+1,j} - V_{i-1,j}}{2\Delta x_i} \qquad (1)$$

$$\left.\frac{\partial^2 V}{\partial x_i^2}\right|_{(i,j)} = \frac{\left.\frac{\partial V}{\partial x_i}\right|_{(i+1,j)} - \left.\frac{\partial V}{\partial x_i}\right|_{(i-1,j)}}{2\Delta x_i} = \frac{V_{i+2,j} - 2V_{i,j} + V_{i-2,j}}{4(\Delta x_i)^2} \qquad (2)$$

$$\left.\frac{\partial^2 V}{\partial x_j \partial x_i}\right|_{(i,j)} = \left.\frac{\partial^2 V}{\partial x_i \partial x_j}\right|_{(i,j)} = \frac{\left.\frac{\partial V}{\partial x_i}\right|_{(i,j+1)} - \left.\frac{\partial V}{\partial x_i}\right|_{(i,j-1)}}{2\Delta x_j}$$
$$= \frac{V_{i+2,j+2} - V_{i-1,j+1} - V_{i+1,j-1} + V_{i-2,j-2}}{4\Delta x_i \Delta x_j} \qquad (3)$$

The variable value in each cell is the averaged value. For the boundary cells,

$$\left.\frac{\partial V}{\partial x_i}\right|_{(1,j)} = \frac{V_{2,j} - V_{0,j}}{\frac{3}{2}\Delta x_i} \qquad (4)$$

$$\left.\frac{\partial^2 V}{\partial x_i^2}\right|_{(1,j)} = \frac{\left.\frac{\partial V}{\partial x_i}\right|_{(2,j)} - \left.\frac{\partial V}{\partial x_i}\right|_{(0,j)}}{\frac{3}{2}\Delta x_i} = \frac{\frac{V_{3,j}-V_{1,j}}{2\Delta x_i} - \frac{V_{1,j}-V_{0,j}}{\frac{1}{2}\Delta x_i}}{\frac{3}{2}\Delta x_i} = \frac{V_{3,j} - 5V_{1,j} + 4V_{0,j}}{3(\Delta x_i)^2} \qquad (5)$$

$$\left.\frac{\partial^2 V}{\partial x_i \partial x_j}\right|_{(1,j)} = \left.\frac{\partial^2 V}{\partial x_j \partial x_i}\right|_{(1,j)} = \frac{\left.\frac{\partial V}{\partial x_j}\right|_{(2,j)} - \left.\frac{\partial V}{\partial x_j}\right|_{(0,j)}}{\frac{3}{2}\Delta x_j} = \frac{V_{2,j+1} - V_{2,j-1} - V_{0,j+1} + V_{0,j-1}}{3\Delta x_i \Delta x_j} \qquad (6)$$

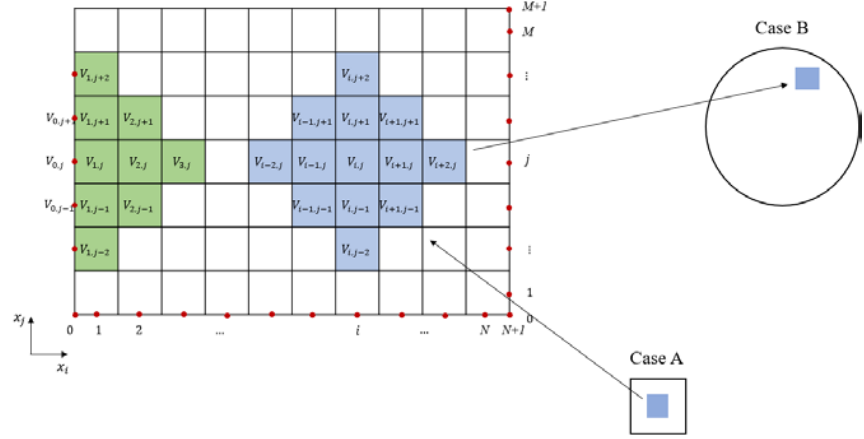

Figure 5. Illustration of How Regional Information is represented by Gradients of Variables in 2D "GELI" Problems

The first part of the local physical parameters is the non-dimensional quantities that represent the local behavior and provide the scalability of physics. This idea came from the early scaling approaches that were applied to develop non-dimensional groups based on facility dimensions and fluid conditions between full-scale facility and scaled test. Another part of local physical parameters as PFs is the parameters involved in the crucial closure correlations for the boundary layer. These parameters contain much model information and are crucial to determining if the PF group provides scalability and superposition of physics. Scalability of physics indicates whether the PF data of the existing case and the target case are similar such that the local physical information of the target case is covered by the existing case, even if these cases are in different global physical conditions. Meanwhile, superposition of physics indicates whether the PFs identified for a simple single phenomenon are still usable for complex coupled physics. The PF group is improvable by adding more relevant PFs if new phenomena are involved, but the computational cost accordingly increases. The determination of optimal PF group is discussed in Step 3.

- **Step 2.2: Collect existing high-fidelity and low-fidelity data**

First is to collect available high-fidelity data that is relevant to the physics involved in the target case. High-fidelity data includes regional data from experimental observation, DNS data, [25] and validated high-resolution numerical results. Since the quantity of high-fidelity data is normally limited, the requirement of "high-fidelity" is flexible, and is determined by the accuracy of expectation on target simulation. For example, if low-fidelity simulation of an NPP containment is executed by coarse-mesh modeling and expected to achieve the accuracy comparable to fine-mesh RANS simulation, then the RANS results can be considered as high-fidelity data. According to the physical conditions of limited high-fidelity data, low-fidelity data is generated using fast-running code with the candidates of mesh sizes and closure models.

- **Step 2.3: Build data warehouse**

Data warehouse includes PF group and simulation errors of local variables. The data of PF group is calculated using low-fidelity simulation data. There are two methods to calculate the error between fine-mesh high-fidelity data and coarse-mesh low-fidelity data: point-to-point and cell-to-cell. The point-to-point method compares the values of local variables at the exact locations existing in both of high-fidelity and low-fidelity data; this method can be applied if both high-fidelity and low-fidelity simulations are using the Finite Element Method (FEM) or the Finite Difference Method (FDM), as shown in Figure 6 (a). Meanwhile, the cell-to-cell method compares the values of local variables in the coarse-mesh cell by averaging and mapping the high-fidelity data from fine cells to coarse ones, as shown in Figure 6 (b). Here the cell-to-cell method is applied for error calculation. Errors of local variables in all coarse cells should be calculated. For example, velocities and temperatures are the main local variables for thermal-hydraulic applications.

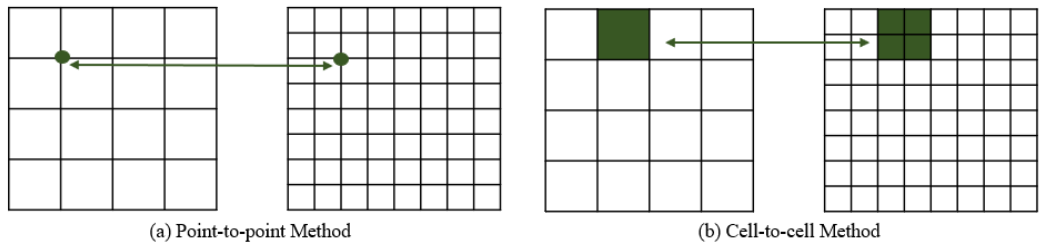

Figure 6. Calculation of Error between Fine-Mesh Data and Coarse-Mesh Data

- **Step 2.4: Determine physics coverage condition of target case**

PF data of the target case should be calculated and compared with the collected data to determine which PCC the target case belongs to. The OMIS framework is only applicable if the target case is located in the GILI condition or the GELI condition; otherwise, the collected data is unusable to cover the local physics of target case. Global parameters defined in Step 1 can be used to specify the global condition while the local condition is qualitatively observed by the t-SNE (t-Distributed Stochastic Neighbor Embedding) method, which is a dimensionality reduction technique for the visualization of high-dimensional datasets. [26]

- **Step 2.5: Design test matrix**

Once the physics coverage condition of the target case is determined, the test matrix should be designed to investigate whether the PF group has the expected predictive capability for the determined physics coverage condition. Physical Feature Coverage (PFC) is the similarity between the training data and testing data that is represented by the coverage (or covered portion) of physical features between training data and testing data. The PFC is quantified using extrapolated distance: further extrapolated distance implies less PFC. The method to calculate extrapolated distance is described in Step 5 since it is also used to guide the construction of optimal training database for target case. The physical condition with similar extrapolated distance should be designed for testing.

- **Step 2.6: Predict error of local variables using machine learning**

Machine learning algorithm is applied to train the error database and obtain the regression function whose inputs are PFs and outputs are the errors of local variables. Currently, multi-layer Feedforward Neural Network (FNN) is identified as the machine learning algorithm for OMIS application. [27]

- **Step 2.7: Evaluate predictive capability on test matrix**

In this step, the originally predicted results for local variables from low-fidelity simulations are adjusted by applying the predicted errors from step 2.6. Then modified values are compared with the high-fidelity data in testing flow. Mean Squared Error (MSE) is used to quantitatively evaluate the predictive capability,

$$MSE_{prediction} = \frac{1}{n}\sum \left(QoI_{HF,i} - QoI_{predicted,i}\right)^2 \qquad (7)$$

Once the comparison with high-fidelity data satisfies the expected accuracy, Step 2 is completed. Otherwise, there are three ways to improve the predictive capability, which are denoted as dashed lines in Figure 3: (1) improving FNN structure, (2) defining more PFs important to the application and (3) collecting more data. Considering their workloads, these three improvements should be performed in the given order. The information flow of Step 2 is described in Figure 7. In training flow, low-fidelity simulations with different mesh sizes and models are performed and compared with high-fidelity data. The local errors ($\varepsilon_i$) of variables between mapped high-fidelity data ($V_{HF,i}$) and low-fidelity simulations ($V_{LF,i}$) are calculated to obtain the error training database. The PF values ($PF_i$) of training flow are obtained using low-fidelity simulation results. The regression error function $\varepsilon = f(PF)$ is obtained by training the database using multi-layer FNN. Then by inserting the new PF values ($PF_j$) of testing flow into the error function, the respective errors ($\varepsilon_j$) can be predicted to modify the low-fidelity simulation results ($V_{LF,j}$). The modified variables ($V_{m,j}$) are compared with the ones from high-fidelity data ($V_{HF,j}$). The predictive capability is tested via validation metric (MSE) to check whether the prediction satisfies the expected accuracy. The accuracy requirements are based on the simulation purpose and limited knowledge of the true physics.

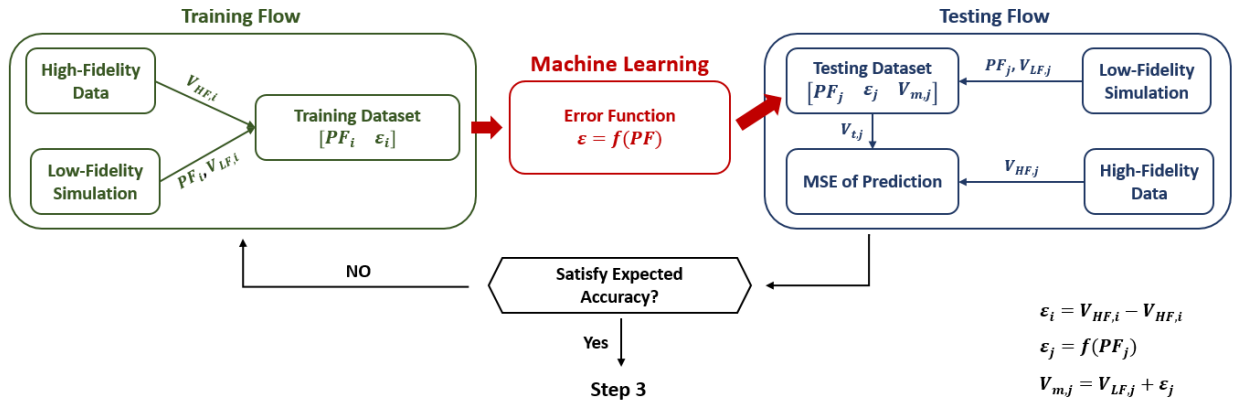

Figure 7. Schematic of OMIS Approach: Training Flow and Testing Flow

### 3.2.3. Step 3: Physical Feature Group Optimization

This step is trying to answer the question: which factors should be considered in the selection of the optimal PF group? According to the definition and classification of PFs discussed in Step 2.1, there are several potential PFs in multi-physics condition. These PFs have different impacts on the responses (errors of local variables). Since training a multi-layer FNN with a huge number of PFs is computationally expensive, it is necessary to identify and rank the importance of each potential PF and select optimal PF group with respect to both PF importance ranking and computation cost for data training.

- **Step 3.1: Identify and rank the importance of all potential physical features**

Importance analysis aims to quantify (1) the change of model output value with respect to the variation of input variables; or (2) the contribution of the uncertainties of input variables to the uncertainty of the model output variable; or (3) the strength of dependence between the model output variable and input variables. [28] The popular importance analysis methods can be divided into two groups: mathematical techniques and statistical techniques. The mathematical techniques include the difference-based methods such as Morris' screening [29], variance-based methods [30], and momentum-based methods [31]. These methods are developed to measure the importance of input variables of models, and most of them need to compute the model response function at prescribed or well-designed points. [32] In the present work, only data (not model) is available to generate input variable (PFs) information, and the inherent correlations between these variables are the key mutual property. Therefore, these traditional sensitivity analysis methods or mathematical techniques are not suitable for identifying the PF importance. Another group, specified as statistical techniques, are designed to explore the variables' importance based on data, including parametric regression and non-parametric regression techniques. These methods are applicable for both computational models and pure data, since the data of input variables can be generated by calling the response function or sampling from a prepared database. Compared to parametric regression methods, non-parametric regression methods do not require a fixed regression model form or an uncorrelated relationship between the input variables. The relationship between PFs and local errors is highly non-linear and affected by the integration of physical models, PF identification, data collection and numerical solvers. There are several popular non-parametric regression techniques such as Gaussian Process Regression [33] and Random Forest Regression (RF regression, or RFR) [34]. In this work, RFR is applied to quantify and rank the PF importance. As a supervised learning algorithm, RFR is much computationally efficient than multi-layer FNN. Compared to traditional statistical methods or other non-parametric regression methods, RFR does not need to assume any formal distributions for the data and can quickly fit highly non-linear interactions even for large problems.

- **Step 3.2: Suggest optimal group of physical features based on the importance ranking**

By performing importance analysis via RFR, the importance of each PF can be quantified and scored. Normally, the scores are in the range from 0 to 10. Based on the scores, the importance of each PF is ranked in three levels: High, Middle and Low (H, M, and L). Different

PF groups can be divided respectively including PFs in H level, H+M level and H+M+L level, where these will be evaluated in the Step 3.3 to determine which PF group is the optimal one for the application. More studies and discussions are needed to determine the adequacy according to the simulation requirements.

- **Step 3.3: Evaluate predictive capability of suggested physical feature group on test matrix**

After the importance identification and ranking, computational cost is saved in the data training for the PF groups only with H level or H+M level. However, uncertainty is also introduced due to reduction of PF dimensionality. The selected PF may be not sufficient to represent the underlying physics, so it is necessary to go back to Step 2.7 and re-test the predictive capability. The selection of an optimal PF group provides the required accuracy with minimal computational cost, where these requirements are defined based on the application. Two metrics should be considered here to finalize the optimal PF group: the MSE of the prediction, and the computational cost for data training.

### 3.2.4. Step 4: Machine Learning Algorithm Determination

This step contains three parts: identify potential FNN candidates, test their predictive capability, and select the optimal one with the consideration of accuracy and computation cost. Several FNN structures with different hidden layers and neuron numbers can be constructed as the potential machine learning method. These FNN candidates should be tested by the test matrix built in Step 2.7 and the optimal FNN structure can be selected based on the MSE of prediction and the computational cost for data training.

### 3.2.5. Step 5: Training Database Construction

Training data is assumed to be sufficient to "cover" the physics in target case; however, some data may be not similar to the target case. It is necessary to select sufficient, applicable datasets as the final training database in order to avoid the huge computational cost on data training.

- **Step 5.1: Define metric to calculate extrapolated distance**

Therefore, we need to answer a question: how to quantitatively measure the similarity of the data in target case and training case? One of the claims is that if target data is more covered or similar to training data, prediction error on target case is smaller. Here, extrapolated distance is firstly defined to (1) determine the similarity between training data and target data and (2) guide the selection/generation of training data source. The goal of extrapolated distance is to measure how far the target point is from the training dataset. The nearest neighbor distance represents the Euclidean distance between the target point and its nearest point in the training dataset. The metrics based on Euclidean distance are easy to compute but very susceptible to noise and memory-consuming since all the points in training dataset are used. Besides, these metrics treat the training data as uncorrelated points and ignore their underlying interactions. There are some promising metrics which are designed memory-efficient by considering the distribution of the training dataset. Mahalanobis distance is defined as the distance between a point ($\boldsymbol{q}$) and the mean of training data ($\boldsymbol{\mu}$) with the covariance matrix ($\boldsymbol{\Sigma}$), which can be expressed as,

$$d_{Ma} = \sqrt{(\boldsymbol{q}-\boldsymbol{\mu})^T \boldsymbol{\Sigma}^{-1}(\boldsymbol{q}-\boldsymbol{\mu})} \quad (8)$$

Mahalanobis distance only considers the statistical parameters like mean and covariance instead of the entire raw data, this makes it more memory efficient. However, the drawback of Mahalanobis distance is its assumption that the training data points should yield a multivariate Gaussian distribution ($\boldsymbol{\mu}, \boldsymbol{\Sigma}$). It is inapplicable to deal with the data from thermal hydraulic simulations, especially for turbulent flows where multi-mode distributions may be common. To overcome this problem, a method called Kernel Density Estimation (KDE) is introduced in this step. KDE is a non-parametric way to estimate the probability density function, which assumes the training data distribution can be approximated as a sum of multivariate Gaussians. One can use a kernel distribution when a parametric distribution cannot properly describe the data, or when one wants to avoid making assumptions about the distribution of the data. KDE can be considered as the probability that the point ($\boldsymbol{q}$) locates in the distribution of training data ($\boldsymbol{p}_i, i = 1, 2, \dots, n$). It is expressed as, [35]

$$p_{KDE} = \frac{1}{n \cdot h_1 h_2 \dots h_d} \sum_{i=1}^{n} \prod_{j=1}^{d} k\left(\frac{q_j - p_{i,j}}{h_j}\right) \quad (9)$$

Where $d$ is the number of variables in $\boldsymbol{q}$ and $\boldsymbol{p}_i$, $k$ is the kernel smoothing function, $h_j$ is the bandwidth for each variable. A multivariate kernel distribution is defined by a smoothing function ($k$) and a bandwidth matrix defined by $H = h_1, h_2, \ldots, h_d$, which control the smoothness of the resulting density curve. Therefore, KDE can be used to measure the distance by estimating the probability of a given point being in a set of training data points. In this step, the KDE distance is standardized as,

$$d_{KDE} = 1 - \frac{p_{KDE}}{p_{KDE} + 0.1} \qquad (10)$$

Before the calculation of the KDE distance, the data of PFs should be normalized in the range [0, 1]. Then the normalized KDE distance will range from 0 to 1. A higher value of the KDE distance means a higher level of extrapolation.

- **Step 5.2: Test the performance of defined metric**

This step is proposed to determine whether the defined metric for extrapolated distance can represent the coverage of training data relative to the target data. In other words, does the prediction error decrease with extrapolated distance? A test matrix can be built with the same training database and different testing data sets. The mean of KDE distance for each test case can be calculated as,

$$D_{KDE} = \frac{1}{n} \sum_{i=1}^{n} d_{KDE,i} \qquad (11)$$

where $d_{KDE,i}$ represents the KDE distance in each local cell of the target case. Then check whether the prediction errors of responses decrease monotonically with the value of $D_{KDE}$. If yes, go to Step 5.3. Otherwise, alternative metric should be explored to assess the coverage.

- **Step 5.3: Select the optimal training database by comparing extrapolated distance**

By comparing the extrapolated distance of each candidate of training database, the optimal one can be selected with the smallest value of $D_{KDE}$.

### 3.2.6. Step 6: Mesh/Model Suggestion

After establishing the predictive capability (Step 2) and selecting the optimal PF group (Step 3), machine learning algorithm (Step 4) and training database (Step 5), the error prediction can be performed for the target case. Since the global QoIs normally have the most concerns on simulation analysis, the criterion of the optimal mesh/model combination is whether it can lead to the least prediction error of the global QoIs. The estimated error ($\varepsilon_{global}$) of global QoIs for different combinations of mesh size and model candidates can be expressed as the average of estimated local errors,

$$\varepsilon_{global} = \frac{1}{n}\sum \varepsilon_{local,i} \qquad (12)$$

Select the one with least estimated error of global QoIs as the "optimal" mesh size and model for the target simulation using the low-fidelity code. The estimation on the error of low-fidelity simulation results are provided.

## 4. CASE STUDY

### 4.1. Problem Statement

The mixed convection case with hot air injection at the bottom of one side wall and a vent on the other side wall was simulated using a GOTHIC 2D model, as shown in Figure 8. The height and length of this cavity are both 1m, while the height of the inlet and vent are both 0.2 m. The target case and the data warehouse are listed in Table 1. The global parameters are defined as below,

$$Gr_i = \frac{g(\rho_w - \rho_i)\rho_f H^3}{\mu^2} \qquad (13)$$

$$Re_i = \frac{U_i \rho_i H}{\mu} \qquad (14)$$

$Gr_i$ of the target case is an extrapolation of the cases in the data warehouse while $Re_i$ of the target case is interpolative. This case study is to investigate the performance of OMIS framework in the extrapolation of high $Gr_i$. By training a DNN, the simulation error prediction and optimal mesh/model selection of the target case will be performed.

Table 1. Target Case and Data Warehouse of Case Study

| Case NO. | | $T_i$ (°C) | $U_i$ (m/s) | $Gr_i$ | $Re_i$ |
|---|---|---|---|---|---|
| | 1 | 30 | 0.1 | 1.124E+09 | 5.863E+03 |
| | 2 | 33 | 0.2 | 1.414E+09 | 1.159E+04 |
| Data Warehouse | 3 | 36 | 0.3 | 1.695E+09 | 1.717E+04 |
| | 4 | 39 | 0.4 | 1.967E+09 | 2.262E+04 |
| | 5 | 42 | 0.1 | 2.231E+09 | 5.585E+03 |

|  | 6 | 45 | 0.2 | 2.486E+09 | 1.103E+04 |
|  | 7 | 48 | 0.3 | 2.733E+09 | 1.634E+04 |
|  | 8 | 51 | 0.4 | 2.971E+09 | 2.152E+04 |
|  | 9 | 54 | 0.1 | 3.201E+09 | 5.312E+03 |
|  | 10 | 57 | 0.2 | 3.424E+09 | 1.049E+04 |
|  | 11 | 60 | 0.3 | 3.638E+09 | 1.554E+04 |
| Target case | | 63 | 0.4 | 3.845E+09 | 2.045E+04 |

\* For each case, one high-fidelity simulation is performed by Star CCM+, four low-fidelity simulations are performed by GOTHIC with different coarse meshes (1/10, 1/15, 1/25, 1/30 m). Each case generates 1850 data points.

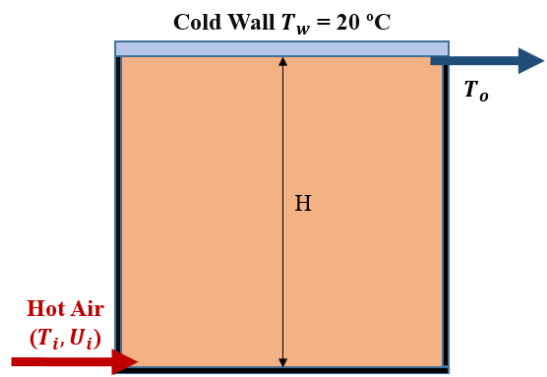

Figure 8. The Illustration of GOTHIC 2D Model for Mixed Convection Case Study

### 4.2. Implementation

#### 4.2.1. Step 1. Simulation Target Analysis

The physics investigated in this case is mixed convection with hot fluid injection and top heat removal. According to the PIRT performed for this analysis [36], mixed convection is mainly dominated by wall friction modeling, turbulence modeling and convection heat transfer modeling, as illustrated in Figure 9.

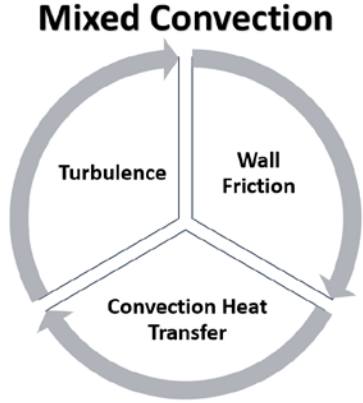

Figure 9. The Illustration of Physics Decomposition for Mixed Convection in Case Study

The turbulence model is the standard two-equation k–ε model with the near-wall treatment in GOTHIC. The convection heat transfer model considers both natural convection and forced convection. The correlations are listed below,

$$h_{nc} = \frac{\lambda}{l} Max(0.54Ra^{0.25}, 0.14Ra^{1/3}) \qquad (15)$$

$$h_{fc} = \frac{\lambda}{l} 0.023 Re^{0.8} Pr^{0.3} \qquad (16)$$

The natural convection model shown in Equation (15) is a mix of two different convection models, which are developed to define a view of flat horizontal surface that is facing down, such as a ceiling in this case study. The forced convection model shown in Equation (16) is developed for pipe flow, which is the only well-defined forced convection model in GOTHIC. The heat transfer coefficient is the maximum of $h_{nc}$ and $h_{fc}$. In this case study, the physical models applied in low-fidelity simulation are fixed, the goal is simplified to predict the simulation error and suggest the optimal mesh size for the target case. Four different mesh sizes are applied for low-fidelity modeling and simulation: 1/10 m, 1/15 m, 1/25 m, and 1/30 m. The global QoI in this case is set as the outlet temperature $T_o$ via the vent, which is directly affected by these key physics.

### 4.2.2. Step 2. Predictive Capability Development

- **Step 2.1: Define potential physical features**

The identified PFs in this case study are marked in red as displayed in Figure 10. Five non-dimensional parameters are defined in this case study: $R$ includes the turbulent information;

$Re$ is defined with the consideration of both $Re$ in free cells and $Re$ in near-wall cells; $Gr$ approximates the ratio of the buoyancy to viscous force acting on a fluid by considering the local density change; $Ri$ expresses the ratio of the buoyancy term to the flow shear term, which represents the importance of natural convection relative to forced convection; $Pr$ reflects the ratio of momentum diffusivity to thermal diffusivity, which depends only on the fluid property and state.

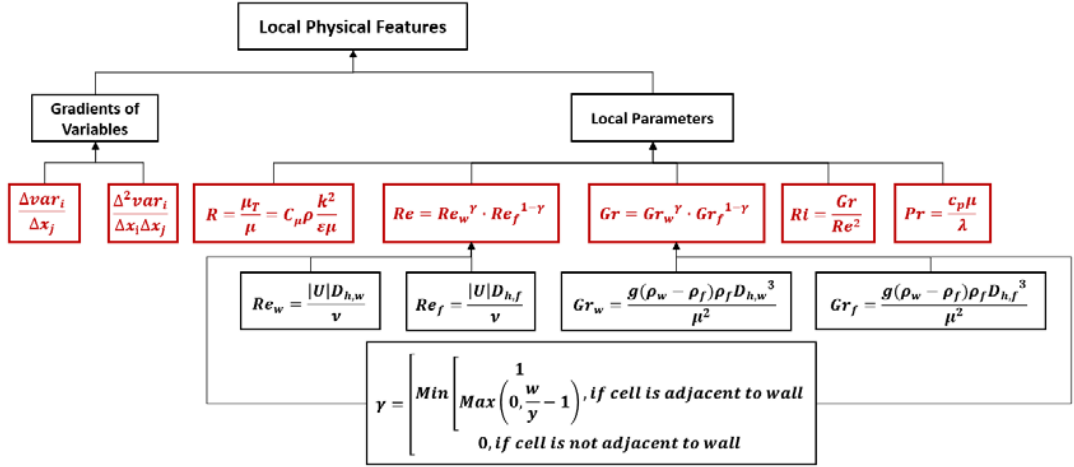

Figure 10. Identification and classification of Physical Features in Case Study

In a pre-test performed before this case study, where only wall friction and turbulence were considered in an adiabatic condition, only $R$ and $Re$ were used. $Gr$, $Ri$ and $Pr$ were added when convection heat transfer became involved.

- **Step 2.2: Collect existing high-fidelity and low-fidelity data**

High-fidelity and low-fidelity data were generated by Star CCM+ with fine mesh and GOTHIC with coarse mesh respectively, as displayed in Figure 11. In this case study, high-fidelity data is generated using 2D RANS model in Star CCM+ with a nodalization of 150x150 in bulk and 600 refinement on top and bottom layer. The refinement on top and bottom is designed to capture the detailed information from injection in the bottom part, and heat removal and venting in the top part. Standard $k$–$\varepsilon$ low-Re model is applied with all y+ wall treatment since (1) this model is robust and easy to implement in small pressure gradient and good for mixing simulation; (2) low Re number approach provides identical coefficients to standard $k$–$\varepsilon$ model and damping functions; (3) all y+ wall treatment is a hybrid treatment that emulates the low y+ wall treatment for fine meshes and the high y+ wall treatment for coarse meshes.

Low-fidelity data for this case study is generated by GOTHIC in four groups with the same closure models and different uniform mesh sizes: 1/10 m, 1/15 m, 1/25 m, and 1/30 m. In Figure 11, the 2D nodalizations in Star CCM+ and GOTHIC (10x10) are displayed with the temperature distribution and horizontal velocity magnitude.

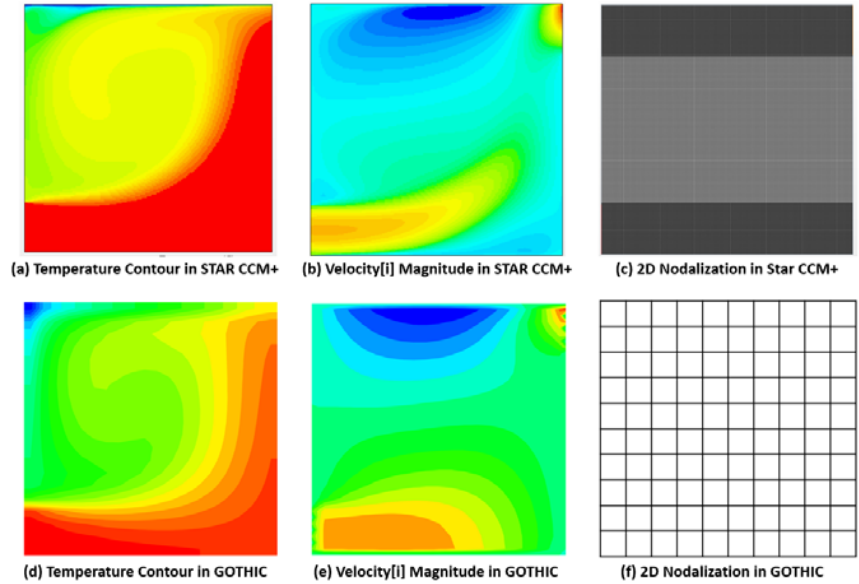

Figure 11. Illustration of 2D GOTHIC Model with Coarse Meshes and 2D Star CCM+ Model with Fine mesh

- **Step 2.3: Build data warehouse**

The PFs are defined in Step 2.1, and the local QoIs are velocities and temperatures $(u, v, T)$. The cell-to-cell method is applied to calculate the errors in this case study. The inputs and outputs of the data-driven error function are listed in Table 2, where variables are $u, v, T, P, k$. The database includes the data from cases 1 to 11 in Table 1.

Table 2. Inputs and Outputs of Data-driven Error Function

|  | Physical Feature | Number |
|---|---|---|
| Inputs | $\frac{\Delta var_i}{\Delta x_j} + \frac{\Delta^2 var_i}{\Delta x_j \Delta x_i}$ | 10 +15 (2D) |
|  | $Re, Gr, Ri, Pr, R$ | 5 |
| Outputs | $\Delta QoI_i = QoI_{i_{HF}} - QoI_{i_{LF}}$ | 3 (2D) |

- **Step 2.4: Determine physics coverage condition of target case**

By using the dimensionality reduction technique t-SNE method, the Physics Feature Coverage (PFC) of the target case can be visualized, as shown in Figure 12. Most of the data points of the target case (red points) are covered by the training data points (black points) in cases 1-11, even though globally, the target case is an extrapolation of the training case. The physics coverage condition of the target case is determined to be the GELI condition. The dataset is reduced from high dimensionality (30 D) to low dimensionality (2D); relative distances among the points are stored and reflected from high dimensionality to low dimensionality.

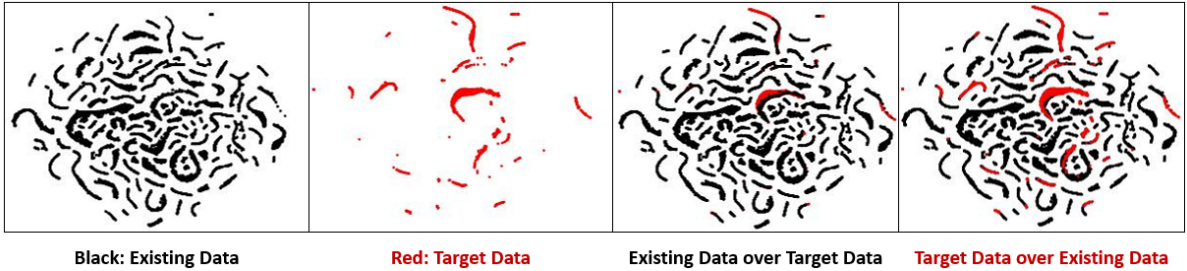

Figure 12. Physical Feature Coverage of Target Case using t-SNE

- **Step 2.5: Design test matrix**

The metric of extrapolated distance is the mean of the KDE distance, which is defined in Equation (11). The physical condition with a similar mean KDE distance should be designated for testing. Different conditions are compared in Table 3. A higher mean of the KDE distance implies less coverage and similarity. The mean of the KDE distance of the target case from case 1-11 is a little smaller than the mean of the KDE distance of case 11 from case 1-7; therefore, here Condition 1 in the test matrix is selected as the test case. If the prediction on case 11 by using case 1-7 as training data is within an acceptable accuracy, the prediction on target case by using case 1-11 is trustworthy since they have similar mean values of KDE distance. Condition 1 is a conservative option as the test case.

Table 3. Test Matrix with Different Training Cases and Testing Cases

| Test Matrix | Testing Case | Training Case | Mean of KDE distance |
|---|---|---|---|
| - | Target case | 1-11 | 0.3354 |
| Condition 1 | 11 | 1-7 | 0.3389 |
| Condition 2 | 11 | 1-8 | 0.3216 |

| Condition 3 | 11 | 1-9 | 0.3067 |
| Condition 4 | 11 | 1-10 | 0.2940 |

- **Step 2.6: Predict error of local variables using machine learning**

In this step, an FNN with 3 Hidden Layers (HLs) and 20 neurons in each hidden layer is applied for data training and prediction on the test case; however, the sensitivity of results to the number of HLs and neurons is assessed later in Step 4.

- **Step 2.7: Evaluate predictive capability on test matrix**

The original low-fidelity simulation results are compared with modified values by machine learning prediction, as shown in Figure 13. The vertical axis is the high-fidelity data. The values of predicted variables (red circles) are quite close to the values from high-fidelity data with small values of MSE. Blue points are the comparison between low-fidelity results and high-fidelity data. The proposed data-driven approach represents good predictive capability and scalability on estimating the local simulation error even for the extrapolation of global physics. The MSEs of predictions are listed in Table 4.

Table 4. MSEs of Predictions for the Test Case in Step 2.7

| Testing Case | Training Case | MSE (u) | MSE (v) | MSE (T) |
| --- | --- | --- | --- | --- |
| 11 | 1-7 | 1.0e-3 | 9.0e-4 | 2.65 |
| Original Low-Fidelity Simulation | | 9.3e-3 | 9.0e-3 | 24.3 |

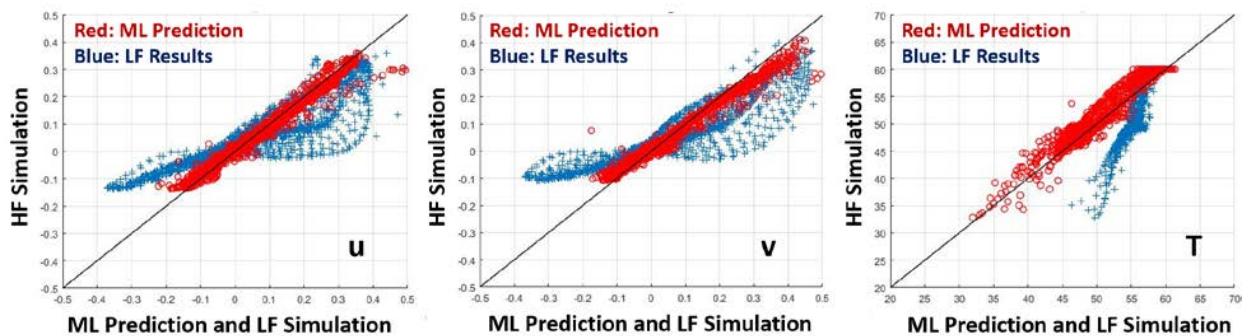

Figure 13. Comparisons between Original Low-Fidelity Simulation Results (LF) and Modified Results from the Machine Learning (ML) Prediction

### 4.2.3. Step 3. Physical Feature Group Optimization

- **Step 3.1: Identify and rank the importance of all potential physical features**

By applying Permutation Variable Importance Measure (PVIM) based on Random Forest Regression (RFR), the importance values of all potential PFs are quantified and ranked, as shown in Figure 14. Higher value implies higher importance. The gradients of velocity, temperature and kinetic energy are more important than the gradients of pressure, since pressure does not change much in the entire cavity. The gradients of pressure are relatively implicit compared with others. All the local physical parameters have great importance.

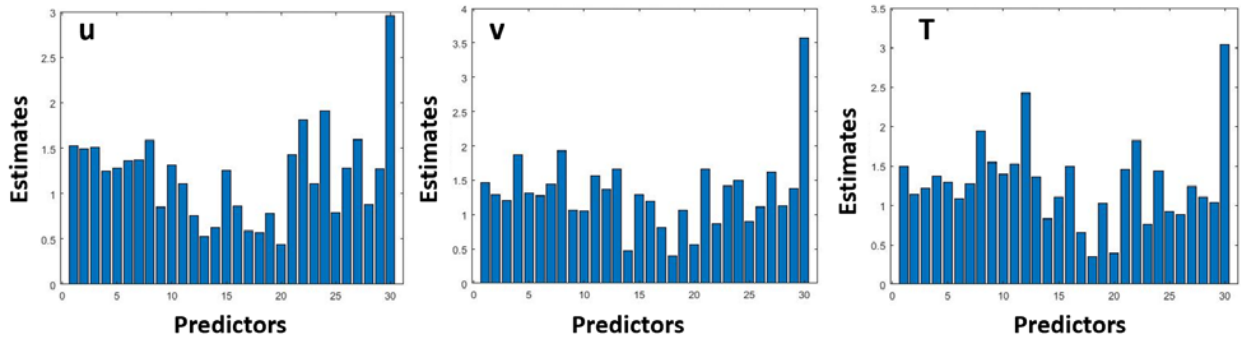

Figure 14. Importance Estimation of PFs on Different Local FOMs using RFR

- **Step 3.2: Suggest optimal group of physical features based on the importance ranking**

According to the importance scores of each PF, the importance of PF can be manually classified into three levels: High, Middle and Low (H, M, and L). Each PF has 3 importance scores. High level means all these three scores of this PF are higher than 1; Low level means all the scores are less than 1; Middle level represents other conditions. The importance classification of each PF is listed in Table 5. Therefore, three different PF groups can be generated respectively including PFs in H level, H+M level and H+M+L level.

Table 5. Importance Classification of Physical Features

| NO. | PF | Importance Level | NO. | PF | Importance Level | NO. | PF | Importance Level |
|---|---|---|---|---|---|---|---|---|
| 1 | $\dfrac{\Delta u}{\Delta x}$ | H | 11 | $\dfrac{\Delta t}{\Delta x}$ | H | 21 | $\dfrac{\Delta k}{\Delta x}$ | H |

| 2 | $\dfrac{\Delta u}{\Delta y}$ | H | 12 | $\dfrac{\Delta t}{\Delta y}$ | M | 22 | $\dfrac{\Delta k}{\Delta y}$ | M |
|---|---|---|---|---|---|---|---|---|
| 3 | $\dfrac{\Delta v}{\Delta x}$ | H | 13 | $\dfrac{\Delta p}{\Delta x}$ | M | 23 | $\dfrac{\Delta^2 k}{\Delta x \Delta x}$ | M |
| 4 | $\dfrac{\Delta v}{\Delta y}$ | H | 14 | $\dfrac{\Delta p}{\Delta y}$ | L | 24 | $\dfrac{\Delta^2 k}{\Delta y \Delta y}$ | H |
| 5 | $\dfrac{\Delta^2 u}{\Delta x \Delta x}$ | H | 15 | $\dfrac{\Delta^2 t}{\Delta x \Delta x}$ | H | 25 | $\dfrac{\Delta^2 k}{\Delta x \Delta y}$ | L |
| 6 | $\dfrac{\Delta^2 u}{\Delta y \Delta y}$ | H | 16 | $\dfrac{\Delta^2 t}{\Delta y \Delta y}$ | M | 26 | $Re$ | M |
| 7 | $\dfrac{\Delta^2 v}{\Delta x \Delta x}$ | H | 17 | $\dfrac{\Delta^2 p}{\Delta x \Delta x}$ | L | 27 | $R$ | H |
| 8 | $\dfrac{\Delta^2 v}{\Delta y \Delta y}$ | H | 18 | $\dfrac{\Delta^2 p}{\Delta y \Delta y}$ | L | 28 | $Gr$ | M |
| 9 | $\dfrac{\Delta^2 u}{\Delta x \Delta y}$ | M | 19 | $\dfrac{\Delta^2 t}{\Delta x \Delta y}$ | M | 29 | $Ri$ | H |
| 10 | $\dfrac{\Delta^2 v}{\Delta x \Delta y}$ | H | 20 | $\dfrac{\Delta^2 p}{\Delta x \Delta y}$ | L | 30 | $Pr$ | H |

* H: all scores > 1.0, L: all scores < 1.0, M: others.

- **Step 3.3: Evaluate predictive capability of suggested physical feature group on test matrix**

Three PF groups with different importance levels are applied and investigated in this step. In Table 6, MSEs of Group 2 (G2) are very close to the MSEs of G3, while the training time is less than half. Therefore, G2 with PFs at H and M levels is used as the optimal PF group in this case study. The physical features in G2 are listed in Table 7. The number of PFs is reduced from 30 to 25. It is expected to reduce more for more complex conditions. It should be noted that more groups can be generated if needed, and predictive capability can be improved using complex FNNs. No matter which PF group is selected, results from the low-fidelity simulation are greatly improved.

Table 6. Predictive Capability of Different PF Groups on Test Case

| PF Group | NO. of PF | MSE (u) | MSE (v) | MSE (T) | Training Time | Testing Case | Training Case | FNN Structure |
|---|---|---|---|---|---|---|---|---|
| G1 (H) | 16 | 6.0e-3 | 4.2e-3 | 3.73 | 1 h | 11 | 1-7 | 3-HL 20-Neuron |
| G2 (H+M) | 25 | 1.1e-3 | 1.1e-3 | 2.69 | 1.5 h | | | |

| G3 (All) | 30 | 1.0e-3 | 9.0e-4 | 2.65 | 3.5 h | | | FNN |
| --- | --- | --- | --- | --- | --- | --- | --- | --- |
| Original Low-Fidelity Simulation | | 9.3e-3 | 9.0e-3 | 24.3 | | | | |

Table 7. The Optimal PF Group for Case Study

| Optimal Physical Features | | Number |
| --- | --- | --- |
| Variable Gradients | $\frac{\Delta u}{\Delta x}$, $\frac{\Delta u}{\Delta y}$, $\frac{\Delta v}{\Delta x}$, $\frac{\Delta v}{\Delta y}$, $\frac{\Delta t}{\Delta x}$, $\frac{\Delta t}{\Delta y}$, $\frac{\Delta p}{\Delta x}$, $\frac{\Delta k}{\Delta x}$, $\frac{\Delta k}{\Delta y}$, $\frac{\Delta^2 u}{\Delta x \Delta x}$, $\frac{\Delta^2 u}{\Delta y \Delta y}$, $\frac{\Delta^2 v}{\Delta x \Delta x}$, $\frac{\Delta^2 v}{\Delta y \Delta y}$, $\frac{\Delta^2 u}{\Delta x \Delta y}$, $\frac{\Delta^2 v}{\Delta x \Delta y}$, $\frac{\Delta^2 t}{\Delta x \Delta x}$, $\frac{\Delta^2 t}{\Delta y \Delta y}$, $\frac{\Delta^2 t}{\Delta x \Delta y}$, $\frac{\Delta^2 k}{\Delta x \Delta x}$, $\frac{\Delta^2 k}{\Delta y \Delta y}$ | 20 |
| Local Parameters | $Re, Gr, Ri, Pr, R$ | 5 |

### 4.2.4. Step 4. Machine Learning Algorithm Determination

In this step, different multi-layer FNN structures are investigated for the test case as displayed in Table 8, which shows that the 4-HL 20-neuron FNN has the most promising performance: higher accuracy and less computational cost. The predictive performance using 4-HL 20-neuron FNN for training and prediction is shown in Figure 15. Compared with the original low-fidelity simulation results (blue points), the values of predicted variables (red circles) are much closer to the values mapped from high-fidelity data with small MSE. Therefore, the 4-HL 20-neuron FNN is selected as the optimal FNN structure and machine learning algorithm in the following steps.

Table 8. Performance of FNN Candidates for Test Case

| # of Hidden Layer | # of Neuron | Training Time | MSE (u) | MSE (v) | MSE (T) | Testing Case | Training Case |
| --- | --- | --- | --- | --- | --- | --- | --- |
| 1 | 20 | 0.5 h | 3.2e-3 | 4.4e-3 | 9.13 | | |
| 2 | 20 | 1 h | 2.4e-3 | 1.3e-3 | 3.02 | | |
| 3 | 20 | 1.5 h | 1.1e-3 | 1.1e-3 | 2.69 | 11 | 1-7 |
| 3 | 30 | 9 h | 1.2e-3 | 7.2e-4 | 1.57 | | |
| 4 | 20 | 6 h | 8.2e-4 | 7.1e-4 | 1.07 | | |
| Original Low-Fidelity Simulation | | | 9.3e-3 | 9.0e-3 | 24.3 | | |

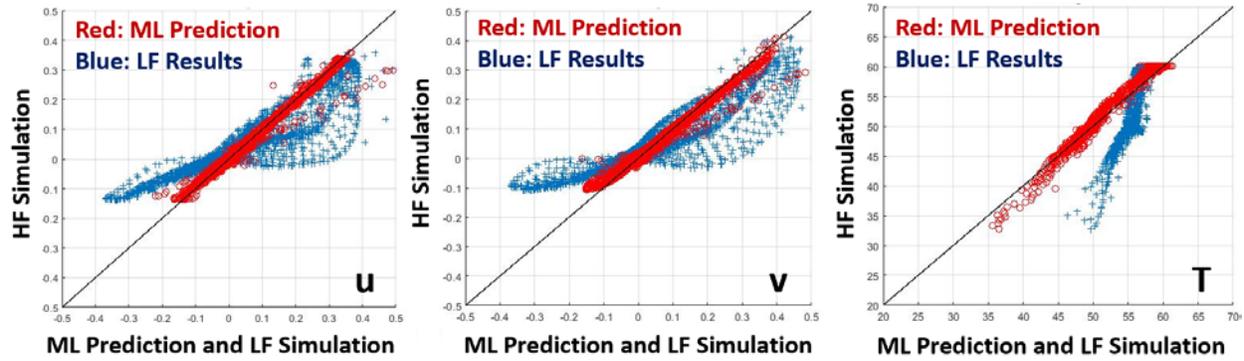

Figure 15. Predictive Performance using 4-HL 20-neuron FNN

### 4.2.5. Step 5. Training Database Construction

- **Step 5.1: Define metric to calculate extrapolated distance**

The KDE distance is applied as the extrapolated distance. One can use a kernel distribution when a parametric distribution cannot properly describe the data, or when one wants to avoid making assumptions about the distribution of the data. KDE can be used to measure the distance by estimating the probability of a given point being located within a set of training data points. KDE distance is standardized as in Equation (10).

- **Step 5.2: Test the performance of defined metric**

In this step, the capability of KDE distance to represent the coverage of training data on target data is evaluated. The mean of the KDE distance for each test case can be calculated in Equation (11). Several tests are performed to explore the relationship between mean of KDE distance and MSEs of prediction, as displayed in Table 9. It seems that there is a nearly positive relationship between mean of KDE distance and MSEs of prediction, as displayed in Figure 16. With higher mean of KDE distance, the MSEs of prediction tends to increase. This relationship is expected to be more distinct when more data are included.

Table 9. Mean of KDE distance and MSEs of Prediction of Tests

| Training Cases | Testing Case | FNN Structure | Mean of KDE distance | MSE (u) | MSE (v) | MSE (T) |
|---|---|---|---|---|---|---|
| 1-7 | 8 | 4-HL 20-Neuron | 0.2282 | 8.9e-5 | 8.54e-5 | 0.20 |
| | 9 | | 0.2493 | 8.5e-4 | 8.6e-4 | 2.01 |

|     | 10 |     | 0.2834 | 1.0e-3  | 9.0e-4  | 2.11 |
|     | 11 |     | 0.3450 | 1.1e-3  | 1.1e-3  | 2.69 |
|     | 9  |     | 0.2442 | 5.2e-4  | 6.5e-4  | 1.46 |
| 1-8 | 10 |     | 0.2732 | 9.36e-4 | 7.53e-4 | 1.75 |
|     | 11 |     | 0.3269 | 1.20e-3 | 1.03e-3 | 1.72 |
| 1-9 | 10 |     | 0.2622 | 8.7e-4  | 9.0e-4  | 2.34 |
|     | 11 |     | 0.3112 | 1.14e-3 | 9.5e-4  | 1.79 |

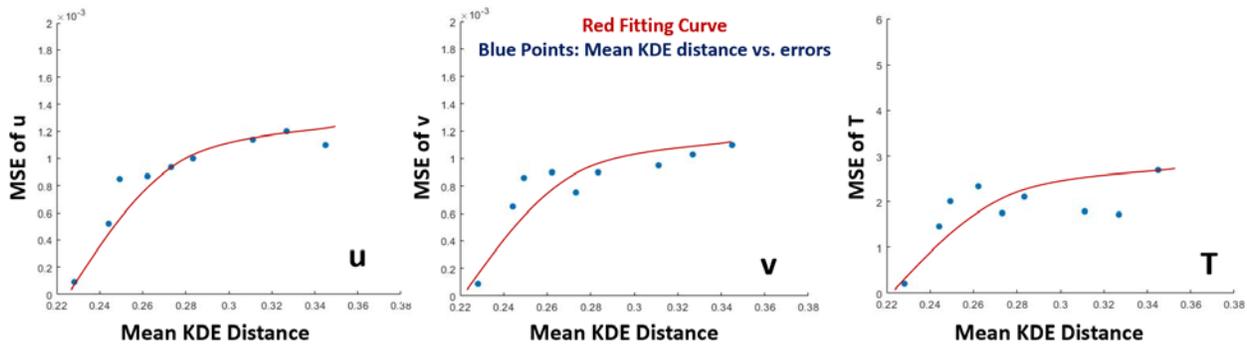

Figure 16. Relationship between Mean of KDE distance and MSEs of Prediction

- **Step 5.3: Select the optimal training database by comparing extrapolated distance**

By comparing the mean of the KDE distance of each candidate in the training database, the optimal one can be selected with the smallest value of KDE distance, as shown in Table 10. According to the values of mean of KDE distance, case 11 has very similar data as the target case. Although the training database with case 3-11 has smaller mean of KDE distance than the one of training database with case 2-11, the latter one is selected as the optimal training database since the prediction error does not change much when the mean of the KDE distance exceeds 0.3, according to Figure 16. In addition, the performance of the multi-layer FNN relies on the size of training database; it needs to include more data to fully capture the underlying information. Based on the FNN performance and the computational cost of FNN training, the training database with case 2-11 is selected as the optimal training database.

Table 10. Mean of KDE distance of Training Database Candidates

| Testing Case | Training Cases | Mean of KDE distance |
| --- | --- | --- |
| Target case | 1-11 (all) | 0.3388 |
| | 1-10 | 0.3542 |
| | 2-11 | 0.3253 |
| | 3-11 | 0.3126 |

### 4.2.6. Step 6. Mesh/model Suggestion

The error prediction of local FOMs can be performed for the target case by using the optimal PF group, the optimal DNN structure, and the optimal training database. In this case study, the global QoI is defined as the outlet temperature. The estimated error of global QoIs ($\varepsilon_{global}$) for different combinations of mesh size candidates and model candidates can be expressed as the average of estimated local error, as in Equation (12). The outlet temperature is calculated as the average shown in Figure 17 for four different mesh sizes. The predicted errors of outlet temperature with different mesh sizes are listed in Table 11. According to the error prediction, GOTHIC simulation with the mesh size as 1/30 m has the least predicted error of outlet temperature. Therefore, 1/30 m is the optimal mesh size for this application and the predicted error of outlet temperature is 0.89.

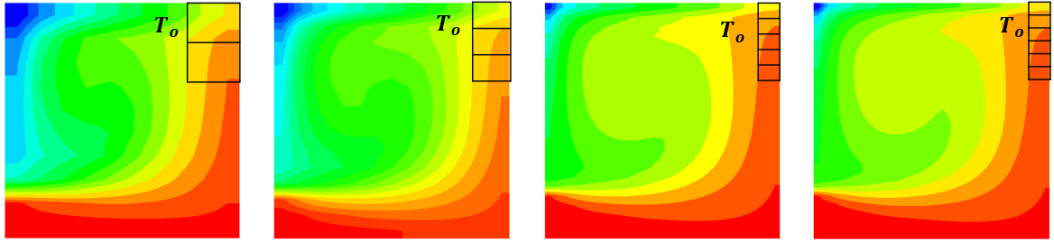

Figure 17. Illustration of Outlet Temperature Calculation in Each Coarse-mesh Simulation

Table 11. Predicted Error of Outlet Temperature with Different Mesh Sizes

| Mesh and Model Candidates | | Predicted Error of $T_o$ |
|---|---|---|
| Model is fixed in this case | 10x10 | 1.88 |
| | 15x15 | 0.96 |
| | 25x25 | 1.74 |
| | 30x30 | 0.89 |

### 4.3. Discussion

After the application of the framework on this case study, we can evaluate:

1. Whether 1/30 m is the optimal mesh size for the target case when the physical model selection is fixed, compared to 1/10 m, 1/15 m and 1/25 m. The comparison between original low-fidelity simulation error and predicted error from OMIS are displayed in Table 12. When the 1/30 m is used as the mesh size, low-fidelity simulation has the least simulation error on the prediction of outlet temperature. Even though OMIS determined that the finest mesh size (1/30 m) tested in this case study was the optimal selection, it does not denote that the coarse-mesh codes "converge" as mesh size goes to zero. Additional runs with finer mesh sizes have been performed, it shows that the simulation error of low-fidelity coarse-mesh codes is not monotonic with mesh sizes.

2. Whether the error prediction on outlet temperature is acceptable. By comparing low-fidelity simulation error and predicted error from OMIS in Table 12, the error of prediction from OMIS is calculated and much smaller than 1%. It is sufficiently accurate and quite acceptable.

Table 12. Comparison of Original Low-Fidelity Simulation Error, Predicted Error by OMIS and Prediction Error of Outlet Temperature

| Mesh Size | $T_{o_{LF}}$ | $T_{o_{HF}}$ | $T_{o_{predicted}}$ | Low-Fidelity Simulation Error | Predicted Error | Relative Prediction Error |
|---|---|---|---|---|---|---|
| 10x10 | 59.11 | 61.08 | 60.99 | 1.97 | 1.88 | 0.15% |
| 15x15 | 60.33 | 61.43 | 61.29 | 1.1 | 0.96 | 0.22% |
| 25x25 | 60.01 | 61.64 | 61.75 | 1.63 | 1.74 | 0.17% |
| 30x30 | 60.74 | 61.68 | 61.63 | 0.94 | 0.89 | 0.08% |

3. Whether low-fidelity simulation can be well corrected by OMIS framework. The corrected results by machine learning are compared with the original low-fidelity simulation as displayed in Figure 18. Low-fidelity simulation is greatly improved by applying OMIS framework.

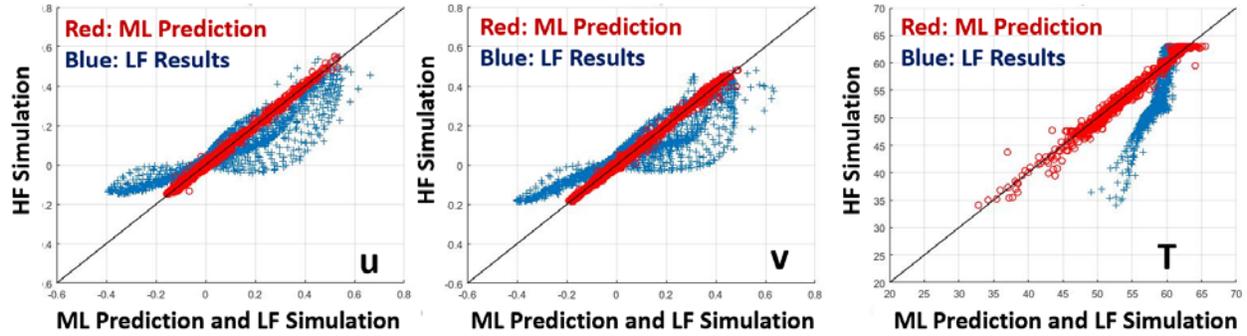

Figure 18. Comparisons between Low-Fidelity Simulation Results (LF) and Corrected Results by OMIS framework for the Simulation of Target Case with 1/30 m as Mesh Size (ML)

## 5. CONCLUSIONS

In this work, a data-driven framework (OMIS) was developed and demonstrated to improve applications of the coarse-mesh codes by predicting their simulation errors and suggesting the optimal mesh size and closure models. The OMIS framework was illustrated based on the mixed convection case study. By learning from massive data instead of human experience, the OMIS framework provides data-driven guidance to help a user improve the modeling and simulation. This modularized procedure is extendible to modeling and simulation using other coarse-mesh codes where mesh size is one of the key model parameters. Scalability of the OMIS framework in the GELI condition is achieved by exploring local physics instead of global physics with the usage of advanced deep learning techniques and statistical approaches. It is expected to improve the scale-distorted approaches that connect scaled data to the real full-scale applications and reduce the uncertainty of scaling. This work also contributes to the development of a data-driven framework for the validation and uncertainty quantification of CFD-like codes. The OMIS framework treats physical models, coarse mesh sizes, and numerical solvers as an integrated model, which can be considered as a surrogate for governing equations and closure correlations. The prediction of simulation error takes all the error sources into account and has promising accuracy even for extrapolated conditions where validation data is not available.

Limitations of the proposed framework exist. The proposed framework has only been demonstrated on a synthetic example in steady state, and the uncertainty quantification of the OMIS framework due to the insufficiency of training data and machine learning application needs more effort.

**ACKNOWLEDGMENTS**

Authors gratefully acknowledge the support by the Idaho National Laboratory's National University Consortium (NUC) program and the INL Laboratory Directed Research & Development (LDRD) program under DOE Idaho Operations Office Contract DE-AC07-05ID14517. The authors are also grateful for partial support from the U.S. Department of Energy's Consolidated Innovative Nuclear Research program via the Integrated Research Project on "Development and Application of a Data-Driven Methodology for Validation of Risk-Informed Safety Margin Characterization Models" under the grant DE-NE0008530. GOTHIC incorporates technology developed for the electric power industry under the sponsorship of EPRI, the Electric Power Research Institute. This work was completed using a GOTHIC license for educational purposes provided by Zachry Nuclear Engineering, Inc.

**APPENDIX A. T-DISTRIBUTED STOCHASTIC NEIGHBOR EMBEDDING**

t-SNE (t-Distributed Stochastic Neighbor Embedding) is a dimensionality reduction technique that visualizes high-dimensional data by giving each data point a location in a two or three-dimensional map. [26] It converts the high-dimensional Euclidean distances between data points into conditional probabilities that represent similarities. Its input typically consists of a collection of $N$ high-dimensional data vectors $X = \{x_1, \ldots, x_N\}$, the pairwise distances $\delta_{ij}^2 = \|x_i - x_j\|^2$ between the high-dimensional data points $x_i$ and $x_j$ are converted into a joint probability distribution $P$ over all pairs of non-identical points. The matrix $P$ has entries as,

$$p_{ij} = \frac{\exp(-\delta_{ij}^2/\sigma)}{\sum_k \sum_{l \neq k} \exp(-\delta_{kl}^2/\sigma)}, \quad for\ \forall i \forall j: i \neq j \tag{17}$$

The similarity of $x_j$ to $x_i$ is the conditional probability, $P_{ij}$ that $x_i$ would pick $x_j$ as its neighbor if neighbors were picked in proportion to their probability density under a Gaussian centered at $x_i$. For nearby data points, $p_{ij}$ is relatively high, whereas for widely separated data points, it will be almost infinitesimal (for reasonable values of the variance of the Gaussian, $\sigma$).

For the low-dimensional counterparts $y_i$ and $y_j$ of the high-dimensional data points $x_i$ and $x_j$, it is possible to compute a similar conditional probability, which is denoted by $q_{ij}$.

$$q_{ij} = \frac{(1 + \|y_i - y_j\|^2)^{-1}}{\sum_k \sum_{l \neq k}(1 + \|y_k - y_l\|^2)^{-1}}, \quad for\ \forall i \forall j: i \neq j \qquad (18)$$

The key property of t-SNE is that, in the low-dimensional map, the similarity between two points is not proportional to a Gaussian density, but to that of a Student-t distribution with a single degree of freedom. By using a heavy-tailed distribution to measure similarities in the low dimensional map, t-SNE allows points that are only slightly similar to be visualized much further apart in the map. This typically leads to very good visualizations. The error between the input similarities $p_{ij}$ and their counterparts in the low-dimensional map $q_{ij}$ is measured by means of the Kullback-Leibler divergence between the distributions P and Q,

$$C(Y) = KL(P\|Q) = \sum_i \sum_{j \neq i} p_{ij} \log \frac{p_{ij}}{q_{ij}} \qquad (19)$$

The asymmetric nature of the Kullback-Leibler divergence leads the cost function to focus on appropriately modeling the large pairwise similarities $p_{ij}$ between the input objects. In other words, similar input objects really need to be close together in the low-dimensional map in order to minimize the cost function $C(Y)$. As the cost function $C(Y)$ is generally non-convex, the minimization of $C(Y)$ is typically performed using a gradient descent method,

$$\frac{\delta C}{\delta y_i} = 4 \sum_j (p_{ij} - q_{ij})(y_i - y_j)(1 + \|y_i - y_j\|^2)^{-1} \qquad (20)$$

**APPENDIX B. IMPORTANCE ANALYSIS BY RANDOM FOREST REGRESSION**

Random Forest Regression (RFR) is an ensemble learning technique that works by constructing a forest of uncorrelated regression trees at training time, and outputting a mean prediction from these individual trees. The training algorithm for random forests applies the general technique of bootstrap aggregating (or bagging). Given a training set $D_N = \{(X_{N \times F}, Y_N)\}$, bagging repeatedly (M times) selects a random subsample ($\Theta_m, m = 1,2,\dots M$) with replacement of the training set and fits trees to these subsamples. N is the number of data points, $F$ is the number of input variables. M regression trees ($\{h(\Theta_m), m = 1,2,\dots M\}$) are trained and built for M times samplings, then provide M times of prediction ($p_{m(x_p)}$) for a new unseen sample ($x_p$). Then the final prediction is made by averaging the M predictions:

$$P = \frac{1}{M} \sum_{m=1}^{M} p_{m(x_p)} \qquad (21)$$

This bootstrapping procedure leads to good predictive performance since it de-correlates these regression trees and decreases the bias of ensemble prediction, by providing different training datasets. In addition, the prediction uncertainty can be estimated as the standard deviation of the predictions from all the individual regression trees:

$$\sigma = \sqrt{\frac{1}{M-1} \sum_{m=1}^{M} \left(p_{m(x_p)} - P\right)^2} \qquad (22)$$

Normally, M is a few hundred to several thousand, depending on the size of the training dataset. Once the regression trees have been built, the importance of variables can be measured by observing the Out-Of-Bag (OOB) error, which is called the Permutation Variable Importance Measure (PVIM) [34]. A set of OOB datasets can be generated as $B_m = D_N - \Theta_m$. The following process describes the estimation of variable importance values by PVIM. Suppose the OOB data can be expressed as $B_m = \{(y_j^m, x_j^m), m = 1,2,\ldots M \text{ and } j = 1,2,\ldots,S\}$, where $S$ is the number of sample points.

1. For the mth tree, the prediction errors on the OOB data before and after randomly permuting the values of the input variable $X_f$ ($f = 1,2,\ldots,F$) are calculated using,

$$MSE_m = \frac{1}{S}\sum_{j=1}^{S}\left(y_j^m - \hat{y}_j^m\right)^2 \text{ and } MSE_{m,f} = \frac{1}{S}\sum_{j=1}^{S}\left(y_j^m - \hat{y}_{j,f}^m\right)^2 \qquad (23)$$

where $\hat{y}_j^m$ and $\hat{y}_{j,f}^m$ are the prediction from the *m*th tree respectively before and after permutation.

2. The difference between two predictions are defined as the value of PVIM for the *m*th tree:

$$PVIM_{m,f} = MSE_{m,f} - MSE_m \qquad (24)$$

3. The overall PVIM of $X_f$ in the OOB data is then calculated as

$$PVIM_f = \frac{\frac{1}{M}\sum_{m=1}^{M} PVIM_{m,f}}{\sigma_f} \quad (25)$$

where $\sigma_f$ is the standard deviation of the differences over the total OOB data. The value of $PVIM_f$ indicates the OOB importance of $X_f$ on the response. In this way, the OOB importance can be measured for each input variable. In the *m*th tree, if $X_f$ is not selected as the splitting variable, then $PVIM_f = 0$. This implies that the interactions between $X_f$ and other variables are considered to measure its contribution on the prediction accuracy. The importance of a variable increases with the value of PVIM.

**APPENDIX C. MULTIVARIATE KERNAL DENSITY ESTIMATION**

A multivariate kernel distribution is a nonparametric representation of the Probability Density Function (PDF) of a random vector. A multivariate kernel distribution is defined by a smoothing function and a bandwidth matrix, which control the smoothness of the resulting density curve. Let $x = (x_1, \dots, x_d)'$ be a *d*-dimensional random vector with a density function *f* and let $y_i = (y_{i1}, \dots, y_{id})'$ be a random sample drawn from *f* for $i = 1, 2, \dots, n$, where *n* is the number of random samples. For any real vectors of *x*, the multivariate kernel density estimator is given by, [35]

$$\hat{f}_H(x) = \frac{1}{n}\sum_{i=1}^{n} K_H(x - y_i), \text{ where } K_H(x) = |H|^{-\frac{1}{2}} K(H^{-\frac{1}{2}}x) \quad (26)$$

where $K(\cdot)$ is the kernel smoothing function, and *H* is the *d*-by-*d* bandwidth matrix. $H^{\frac{1}{2}}$ is a square diagonal matrix with the elements of vector $(h_1, \dots, h_d)$ on the main diagonal. $K(x)$ takes the product form $K(x) = k(x_1)k(x_2)\cdots k(x_d)$, where $k(\cdot)$ is a one-dimensional kernel smoothing function. Then, the multivariate kernel density estimator becomes,

$$\hat{f}_H(x) = \frac{1}{nh_1\cdots h_d}\sum_{i=1}^{n} K\left(\frac{x_1 - y_{i1}}{h_1}, \dots, \frac{x_d - y_{id}}{h_d}\right) = \frac{1}{nh_1\cdots h_d}\sum_{i=1}^{n}\prod_{j=1}^{d} k\left(\frac{x_j - y_{ij}}{h_j}\right) \quad (27)$$